%% file: Ringberg.tex
\documentstyle[times,11pt,paspconf,psfig]{article}
\input{macros_ringberg}

\begin{document}

\title{The Large-Scale Distribution and Motions of Older Stars in Orion}

\author{Anthony G.~A. Brown\altaffilmark{1}}
\affil{Sterrewacht Leiden \\
       P.O. Box 9513, 2300 RA Leiden, The Netherlands \\
       brown@strw.leidenuniv.nl} 

\author{Frederick M. Walter}
\affil{Department of Physics and Astronomy \\
       SUNY Stony Brook, NY 11794-3800, USA \\
       fwalter@astro.sunysb.edu}

\author{Adriaan Blaauw}
\affil{Kapteyn Astronomical Institute \\
       P.O.~Box 800, 9700 AV Groningen, The Netherlands \\
       A.Blaauw@astro.rug.nl}

\altaffiltext{1}{Present address: Instituto de Astronom\'\i a, U.N.A.M.,
P.O. Box 877, Ensenada, 22800 Baja California, Mexico;
brown@bufadora.astrosen.unam.mx}

\begin{abstract}

  We review the current knowledge of the population of `older' stars
  in the Orion OB1 association, specifically those in subgroups 1a and
  1b. We briefly outline the history of the subject and then
  continue with a summary of the present state of knowledge of the
  early-type stars in Orion OB1. New results from the Hipparcos
  parallaxes and proper motions will be presented. The main result
  is that subgroup 1a is located at about 330\,pc from the Sun, much
  closer than the previously determined distance, and about 100\,pc
  distant from the other subgroups of the association and the Orion
  molecular clouds. Unfortunately, due to the unfavorable kinematics
  of the association with respect to the Galactic background,
  Hipparcos proper motions do not allow a clear kinematic separation
  of the association from the field. For this purpose accurate and
  homogeneous radial velocities are needed. Traditionally, the massive~O 
  and~B stars have received most of the attention in the studies of
  OB associations.  However, we will present results showing that
  significant numbers of low-mass stars are associated with Orion OB1.
  Unbiased, optically complete, spectroscopic and photometric surveys
  of areas within subgroups 1a and 1b have the potential to determine
  the complete low-mass stellar population, down to the brown dwarf
  limit.  This will provide much insight into the overall initial mass
  function and studies of the kinematics of the low-mass stars will
  yield insights into the dispersal of the association.

\end{abstract}

\setcounter{footnote}{1}
\keywords{OB associations, star formation, initial mass function}

\section{Introduction}

Ever since spectral classifications for the bright stars became
available it was evident that~O and~B stars are not distributed
randomly on the sky, but instead are concentrated in loose groups
(Blaauw 1964 and references therein). Ambartsumian (1947) found that
the stellar mass density in these groups, which were subsequently
called OB associations, is usually less than 0.1\Msolar\,pc$^{-3}$.
Bok (1934) had already shown that such
low-density stellar groups are unstable against Galactic tidal forces,
so that the observed OB associations must be young, a conclusion
supported by the ages derived from Hertzsprung-Russell
diagrams. These groups are prime sites for the study of star formation
processes and of the interaction of early-type stars with the
interstellar medium (see, \eg{} Blaauw 1964, 1991 for reviews).
Detailed knowledge of the stellar content and structure of OB
associations allows us to address fundamental questions on the
formation of stars in giant molecular clouds. What is the initial mass
function? What are the characteristics of the initial binary
population?  What is the star formation efficiency? Do all stars in a
group form at the same time?  What process causes the distinction
between the formation of bound open clusters and unbound associations?
How is angular momentum redistributed during star formation?

The study of OB associations is thus motivated primarily by the fact
that they form the fossil record of star formation processes in giant
molecular clouds, but is also important in the context of the
evolution of the Galaxy. The kinematics of the nearest OB associations
provides detailed insight into the properties and origin of the Gould
Belt system (\eg{} Blaauw 1991; Elmegreen 1992), which is an example
of the star complexes discussed in the chapter by Elmegreen \& Efremov 
in this volume.

In this context, the Orion complex has received a lot of attention. It
is the nearest giant molecular cloud complex and a site of active star
formation, including the formation of high-mass (O and B) stars. All
stages of the star formation process can be found here, from deeply
embedded protoclusters to fully exposed OB associations.  The
different modes of star formation occurring in these clouds
(clustered, distributed, isolated) allow us to learn more about the
influence of the environment on the star formation process.
Furthermore, the Orion region forms an excellent nearby example of the
effects of early-type stars on the interstellar medium. The
Orion/Eridanus Bubble, surrounding the association and visible in \HI,
H$\alpha$ and soft X-rays is being blown by the supernovae and stellar
winds from the evolved~O and~B stars in the Orion OB1 association
(see chapters by Bally, Theil, \& Sutherland, and Heiles in this
volume). And last but not least, the Orion molecular cloud complex is 
a prime site for studies of interstellar chemistry.

In this contribution the emphasis will be on the older stellar
population of the Orion complex, namely the already exposed subgroups
of the Orion OB1 association. We shall concentrate mostly on the
subgroups 1a and 1b; the other subgroups (1c, around Orion's Sword,
and 1d, containing the Orion Nebula Cluster) are discussed in the
chapters by Allen \& Hillenbrand and McCaughrean \etal{}
in this volume.

We start with a short historical outline of the studies of the Orion
OB1 association, discussing the division into subgroups and
summarizing the work done on the stellar content up to the extensive
photometric study by Warren \& Hesser (1977a, 1977b, 1978). We then
summarize recent work on the massive~O and~B stars 
(mass $\simgreat 4$\Msolar) of the OB association.  This includes 
a discussion of the
Hipparcos proper motions and parallaxes and their implications for the
distance to the Orion complex.

Much work on nearby OB associations has gone into study of the high-mass 
O~and B~stars. For these stars, proper motions from large scale
surveys were available for membership studies, and the contamination
from the Galactic background is minimized in this spectral type range.
However, recent work by, \eg{} Walter \etal{} (1994) on the Upper
Scorpius association (part of Sco OB2; see, \eg{} de\,Geus, de\,Zeeuw,
\& Lub 1989) showed that it contains a field star mass function 
extending down to
at least 0.3\Msolar, with about 2000 members less massive than
about 1\Msolar. We shall also discuss recent work on the region
surrounding the star $\sigma$\,Orionis, which shows that there is a
clustering of low-mass stars around this star. It may possibly be an
older analog of the Orion Nebula Cluster (ONC). The implications for
the initial mass function in Orion will be discussed.

\section{Short Historical Overview}
Much of the material in this section is derived from the review by
Blaauw (1964) and the work by Warren \& Hesser (1977a, 1978). We refer
to these papers for many more details on the early investigations into
the stellar content of the Orion OB1 association. We note that the
overview that follows is not complete and certainly biased towards the
studies of the~O and~B stars in Orion.

The earliest large-scale investigations into the stellar content of
the Orion OB association were those by Shahovskoj (1957), Parenago
(1953, 1954) and Sharpless (1952, 1954, 1962). These studies
concentrated mainly on the stars in the Belt and Sword regions of the
Orion constellation and all concerned spectroscopic and photometric
data. A clear division of the association in subgroups was not yet
established. It was Blaauw (1964) who suggested the division of the
Orion association into four subgroups: 1a, which contains the stars to
the northwest of the Belt stars; 1b, containing the group of stars
located around the Belt (including the Belt stars themselves); 1c, in
which the stars around the Sword are included; and 1d, which contains
the stars in and close to the Orion Nebula (including the Trapezium
stars). The figure of the subgroups shown by Blaauw (1964) shows an
increasing concentration of the subgroup members going from 1a to 1c.
Assuming that the subgroups are unbound and expanding, this suggests a
sequence of decreasing ages, which was confirmed by the studies of the
HR diagrams at that time.

The studies above were followed by the massive photometric
investigation by Warren \& Hesser (1977a, 1977b, 1978). They presented
$uvby\beta$ and $U\!BV$ photometry for 526 stars in the Orion region.
They analyzed the data in terms of reddening and $M_V$ determinations,
possible correlations with stellar axial rotation, and effects on the
photometry caused by anomalous extinction, possibly due to
circumstellar material associated with pre-main sequence stars.
They subsequently determined membership for each subgroup based on the
photometric parallax of the stars. The spatial distribution and the
ages of the subgroups were then studied based on the membership
determinations.

The results of the studies above can be summarized briefly as follows.
The ages of the Orion OB1 subgroups from studies of colour-magnitude
diagrams were listed by Blaauw (1964) as 12, 8, 6, and about 4\,Myr,
going from 1a to 1d. Warren \& Hesser (1978) found the ages to be
7.9, 5.1, 3.7, and $<0.5$\,Myr. Various methods for determining
the ages have been brought to bear and an extensive listing up to 1978
is provided by Warren \& Hesser (1978). The distance to the Orion OB1
association is listed by Blaauw (1964) as 460\,pc. Warren \& Hesser
(1978) determined distances to the subgroups separately and found
distances of 400, 430, 430, and 480\,pc, going from 1a to 1d.  These
authors also list various other distance determinations.

Proper motion studies have focused mainly on the ONC\@.  The earliest of
these studies (Parenago 1954; Strand 1958) showed the Trapezium
Cluster to be unstable and its expansion age to be about 0.3\,Myr.
Investigations of the motions in the more dispersed subgroups were
carried out by Lesh (1968), who derived an expansion age for subgroup
1a of 4.5\,Myr, and Blaauw (1961), who derived kinematic ages of
2.2--4.9\,Myr for the three runaway stars, AE\,Aur, $\mu$\,Col, and
53\,Ari, probably originating from subgroup 1b.  Adding the ages of the
progenitors of the supernovae responsible for the runaways, led to an
age of subgroup 1b in rough agreement with the photometric age (at
that time) of 5--8\,Myr. Proper motion studies are difficult to carry
out in the Orion region due to the crowding of the stars. Moreover, the
interpretation of the proper motions in the more dispersed subgroups
is not straightforward, because the motion of the association is
directed mostly radially away from the Sun, making it difficult to
detect a common space motion for the members of Orion OB1.

Finally, we comment on the division of Orion OB1 into subgroups. In
several studies, subgroups 1b, 1c and 1d were subdivided further.
Warren \& Hesser (1978) split subgroup 1b into three parts because
Hardie, Heiser, \& Tolbert (1964) and Crawford \& Barnes (1966) found
that the distance of the Belt stars increases from west to east.
However, Brown, de\,Geus, \& de\,Zeeuw (1994) found no significant
differences in their mean distances, and no trend with right ascension
for the 1b stars as claimed by Warren \& Hesser. Morgan \& Lod\' en
(1966) and Walker (1969) had divided 1c into several smaller subgroupings
located close to the Orion Nebula, but Warren \& Hesser found no
evolutionary differences between these groups. Therefore, most
recently, Brown \etal{} (1994) decided to treat subgroups 1b and 1c as
a whole. However, as we shall discuss further on, the issue of the
exact division into subgroups is not yet settled.

\section{Recent Work on the Massive Stars of Orion OB1}

The interstellar medium near Orion OB1 contains several large scale
features, including H$\alpha$ emission extending to Eridanus, partly
observable as Barnard's Loop, and a hole in the \HI{} distribution, which
is surrounded by expanding shells (Goudis 1982). Reynolds \& Ogden
(1979) and Cowie, Songaila, \& York (1979) argued that the coherent gas 
motions in Orion are the result of a series of supernova events which took
place up to 4\,Myr ago, but they ignored the effects of
stellar winds. In the past twenty years a wealth of new data has been
gathered on the large scale interstellar medium in Orion, through
surveys in $^{12}$CO (Maddalena \etal{} 1986), $^{13}$CO (Bally \etal{}
1987), CS (Lada, Bally, \& Stark 1991), the far-infrared (IRAS sky
survey), and \HI{} (Chromey, Elmegreen, \& Elmegreen 1989; Green 1991;
Green \& Padman 1993; Hartmann \& Burton 1997). Much of the work on
the ISM around Orion was reviewed by Genzel \& Stutzki (1989). During
the same period, the theory of stellar winds has been developed to the
extent that their impact on the surrounding medium can be readily
estimated (Kudritzki \etal{} 1989; McCray \& Kafatos 1987; de\,Geus
1991, 1992). 

As a consequence of these extensive studies, Brown \etal{} (1994) 
decided to carry out a new investigation of the stellar content of the 
Orion OB1 association in conjunction with a study of the impact of the 
early-type stars on the surrounding ISM\@.
These authors studied a sample of stars in Orion OB1 that were
included in a 1982 Hipparcos proposal to observe all OB associations
within 800\,pc from the Sun. The stars were studied with the $V\!BLUW$
Walraven photometric system (Lub \& Pel 1977). Physical parameters of
the stars were derived and membership (based on photometric
distances) of the subgroups was determined. The distances to the
subgroups of Orion OB1 as determined by Brown \etal{} (1994) are
smaller than distances derived previously. A distance of 380\,pc was
derived for subgroup 1a; 360\,pc for 1b; and 400\,pc for 1c. The distance to
1d (including the Trapezium) could not be determined reliably by Brown
\etal{} (1994) due to the nebulosity in that region and insufficient stars. 
The smaller distances to the subgroups of Orion OB1 were also found by 
Anthony-Twarog (1982), who reanalyzed the data of Warren \& Hesser with 
a revised calibration of the $uvby\beta$ system.

Brown \etal{} (1994) also derived ages for the subgroups in Orion OB1
by isochrone fitting in the $\log g$--$\log T_{\rm eff}$ plane. The
ages found were $11.4\pm 1.9$\,Myr for 1a, $1.7\pm 1.1$\,Myr for 1b, and
$4.6\pm 2$\,Myr for 1c, and less than 1\,Myr for subgroup 1d. These
results imply that there is not a sequence of decreasing ages going
from 1a to 1d, but that 1b is a young subgroup in between the older
subgroups 1a and 1c. This result was found previously by de\,Zeeuw \&
Brand (1985), and Brown \etal{} (1994) discussed other supporting evidence 
for 1b being younger than 1c. One of the arguments in favor of a sequence
of ages has been the increasing degree of concentration in the early-type 
stars from 1a to 1c. However, as we shall discuss below, the
structure of 1b may be more complicated. The concentration of low-mass
stars around $\sigma$\,Ori suggests that that system is an older analog
of the Trapezium (Section~6). This implies that one cannot treat what is
considered to be subgroup 1b as a whole. Doing so may lead to wrong
inferences about the degree of concentration of the group of early-type stars.

Brown \etal{} (1994) also derived the initial mass function (IMF) for
subgroups 1a, 1b, and 1c. They determined the masses from the surface
gravity, luminosity, and effective temperatures of the stars, and
carefully corrected for the presence of binaries. It was assumed that
the present-day mass function for the interval of masses where the
stars are on the main-sequence is a good approximation of the IMF\@.
The IMF was found to be a single power law: 
$\xi(\log M) {\rm d}\log M = AM^{-B} {\rm d}\log M$, 
where $B=1.7\pm0.2$ for all three subgroups.
The mass-ranges over which the IMF was determined were 4--15\Msolar,
4--120\Msolar, and 7--36\Msolar{} for subgroups 1a, 1b,
and 1c respectively, the limits being set by the age of the subgroup
and the completeness of the observations. Previously, Claudius \&
Grosb\o l (1980) had found a value for $B$ of $1.9\pm0.2$.

Taking into account the IMF and the ages of the subgroups, the
mechanical energy output in the form of stellar winds and supernovae
over the lifetime of the association was estimated by Brown \etal{}
(1994). The total energy output of the association over its lifetime
is of the order of $10^{52}$ ergs. This energy is enough to explain the
observed Orion/Eridanus \HI{} shell surrounding the Orion OB1
association. More details on the Orion/Eridanus Bubble can be found in
the chapter by Bally \etal{} and Heiles in this volume, and in 
Burrows \etal{} (1993) and Brown, Hartmann, \& Burton (1995).

Modern proper motion studies of the ONC include those by Jones \&
Walker (1988), van Altena \etal{} (1988) and McNamara \etal{} (1989).
The results of these studies are summarized by Hillenbrand (1997) and
Hillenbrand \& Hartmann (1998). Most recently, the proper motions in
the ONC were studied by Tian \etal{} (1996). Their results generally
agree with those of the three studies above. Assuming a distance of
470\,pc to the ONC, Tian \etal{} (1996) derive an upper limit on the velocity
dispersion of the ONC of $\sim 2$\kmpers, and they conclude that
the ONC is an unbound system. The most recent large-scale study of
proper motions in Orion is that by Smart (1993), who used photographic
plates to study the area of Orion OB1b and 1a.

Abundance patterns in stars of Orion OB1 have recently been studied in
a series of papers by Cunha \& Lambert (1992, 1994), Cunha, Smith, \&
Lambert (1995), and Cunha \etal{} (1997). The abundance analysis of
these authors shows that the stars in Orion OB1, in common with
the Orion Nebula \HII{} region, are underabundant in oxygen with respect 
to the Sun. The lowest abundances are found in subgroups 1a and 1b. The 
Trapezium stars and some stars of subgroup 1c seem to have abundances that are
up to 40\% higher than those in subgroups 1a and 1b (although
still subsolar). Cunha \& Lambert (1992) suggest that this is due to
enrichment of the interstellar gas by mixing of supernovae ejecta from
subgroup 1c with the gas that subsequently collapsed to form the
Trapezium Cluster. This enrichment scenario is confirmed by the fact
that Cunha \& Lambert (1994) observe no abundance variations for C, N,
and Fe, but do observe the same abundance variations for Si as for
oxygen, as one would predict for cloud material enriched by Type~II
supernova ejecta.  Supernovae must have occurred in Orion OB1 in the
past as evidenced by the presence of the Orion/Eridanus Bubble. Brown
\etal{} (1994) estimated that 1 to 2 supernovae have occurred in
subgroup 1c.

These results suggest a sequence of star formation from the older 1a,
1b, and 1c subgroups to the young ONC\@.  However, this does not imply
that star formation throughout the Orion molecular clouds followed the
formation of the older subgroups of Orion OB1. As discussed by
Hillenbrand (1997), there is a wide range of ages of young stars in the
molecular cloud complex, with most of the low-mass stars in the L\,1641
cloud being older than the stars in 1c and 1d. We shall return to this
point after discussion of the Hipparcos parallaxes of the stars in
Orion OB1.

\section{Hipparcos Results on Orion OB1}
As mentioned above, the stars in the Orion OB1 association were
included in a 1982 proposal to observe all OB associations within 800\,pc 
from the Sun with the Hipparcos satellite. The aim of this project,
called SPECTER, is to carry out a comprehensive census of the stellar
content of nearby OB associations. Detailed knowledge of the stellar
content of nearby associations allows one to address fundamental
questions on the formation of stars in giant molecular clouds, and
comparisons between the different associations may lead to further
insight into the question of whether the outcome of the star formation
process is universal or depends on local conditions. Project SPECTER
was described most recently by de\,Zeeuw, Brown, \& Verschueren (1994).

The original proposal asked for observation of all~O and~B stars and
later-type stars within certain magnitude ranges, in the region around
Orion OB1 defined by the Galactic coordinate limits $196^\circ\le
\ell\le 217^\circ$ and $-27^\circ\le b\le -12^\circ$. Within this
region there are 1142 stars in the Hipparcos Catalogue (ESA 1997; see
Perryman \etal{} 1997 for a summary of the catalogue contents) of
which 294 are O~and B~stars, 391 are A~stars, 208 are F~stars, and the
rest are of type~G or later. Part of the effort for project SPECTER
involved the development of new methods for identifying OB
associations in the Hipparcos data. Although these groups are unbound,
their expansion velocities are only a few \kmpers{} (\eg{} Mathieu
1986; Blaauw 1991), so that the common space motion is perceived as a
motion of the members towards a convergent point on the sky (\eg{}
Blaauw 1946; Bertiau 1958). Two methods that make use of the common
space motion of stars in association were developed. One is an
improved version of the convergent point method employed by Jones
(1971), and the other a newly developed method which uses both the
parallaxes and the proper motions from Hipparcos. These methods are
described in more detail by de Bruijne \etal{} (1997) and preliminary
results on the nearby associations are presented by de\,Zeeuw \etal{} 
(1997) and Hoogerwerf \etal{} (1997). We note that these methods of
membership determination are also strictly applicable in the case of
associations in a state of expansion.

\begin{figure}[t]
\centerline{\psfig{figure=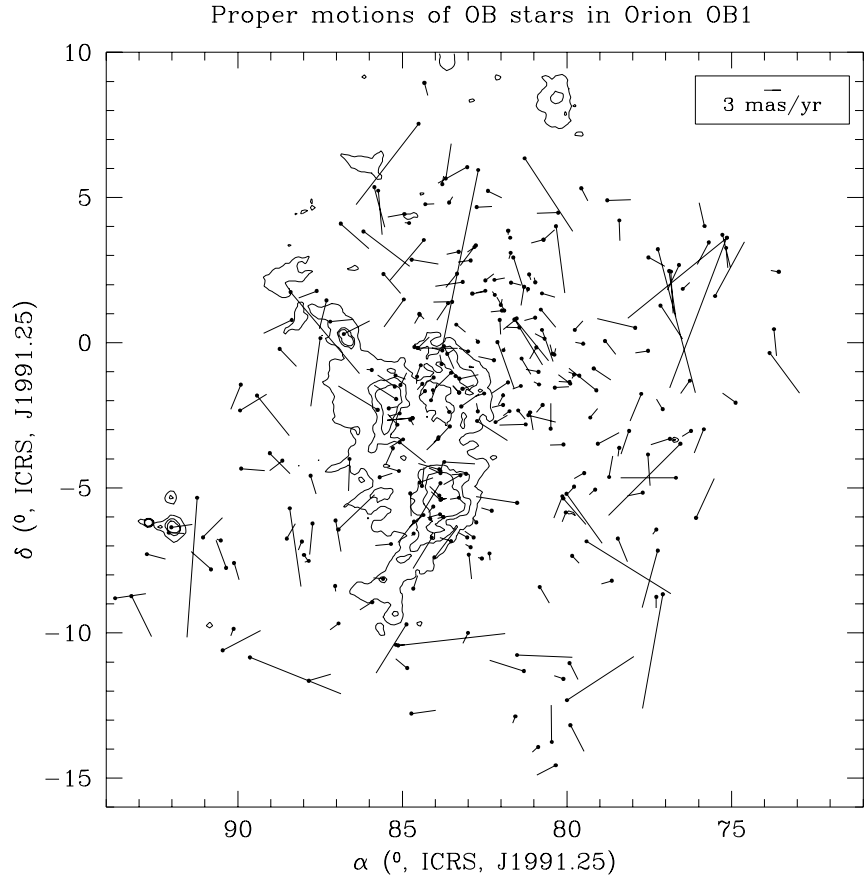,width=12cm}}
\caption{Proper motions of the O and B stars in the region of the sky
  around Orion OB1. The proper motions are indicated as vectors
  normalized to 3\,mas/yr. The contours are the IRAS 100\micron{}
  skyflux and they outline the location of the molecular clouds. Note
  the absence of a clear pattern of common proper motions near the
  Orion OB1 subgroups and the larger proper motions towards the edges
  of the field.}\label{fig-orion_pm}
\end{figure}

\begin{figure}[t]
\centerline{\psfig{figure=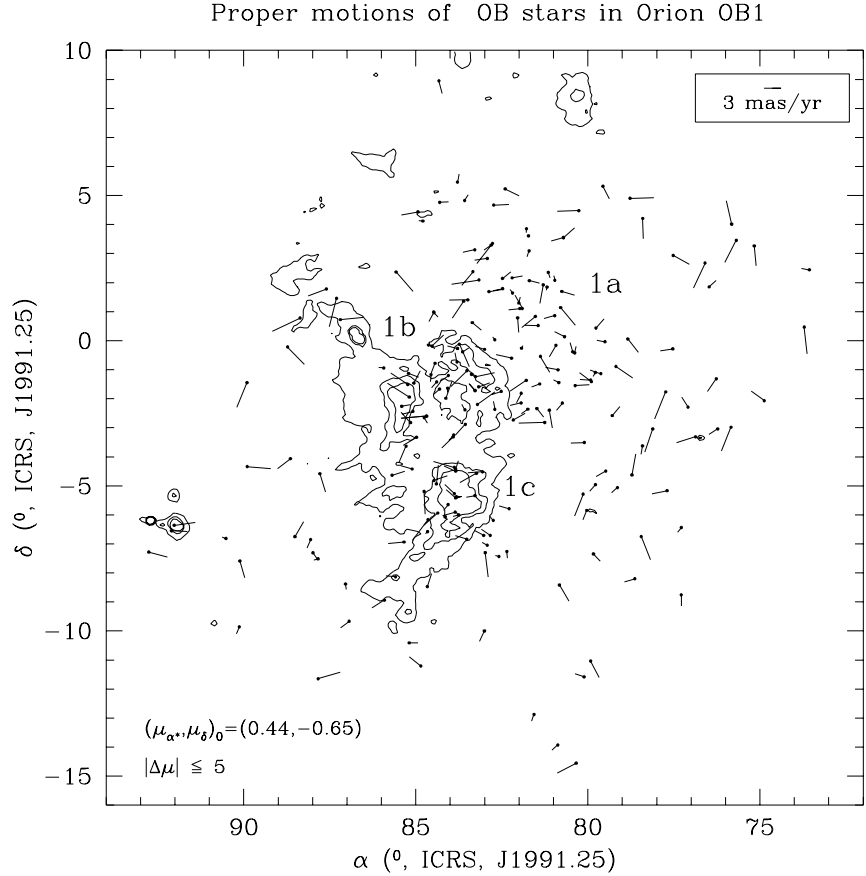,width=12cm}}
\caption{Same as Figure~\ref{fig-orion_pm}, but after selection of the
  stars according to their proper motions. Note that now there is a
  clear concentration of stars towards the known locations of the
  Orion OB1 subgroups. This indicates that the rough selection process
  indeed filters out non-members of the
  association.}\label{fig-orion_pm2}
\end{figure}

Unfortunately in the case of Orion OB1 the motion of the association
is mostly directed radially away from the Sun. This makes it very hard
to detect the association with proper motion studies. This is
illustrated in Figure~\ref{fig-orion_pm}. The figure shows the
Hipparcos proper motions for all the O~and B~stars in the region of
Orion. No clear pattern of motions emerges in the region of the Orion
OB1 subgroups.  However it is clear that most of the large proper
motions occur towards the edges of the field. A rough selection of
Orion members can be made by demanding that:
\begin{equation}
(\mu_\alpha\cos\delta-0.44)^2+(\mu_\delta+0.65)^2\le 25 ~,
\end{equation}
where the units are milli-arcsec per year (mas/yr). The list of stars
obtained largely overlaps with the photometrically determined list of
members from Brown \etal{} (1994). The resulting distribution of stars
is shown in Figure~\ref{fig-orion_pm2}. There is a clear concentration
of stars towards the well known subgroups of Orion OB1.  Among the OB
stars along the edges of the field that remain after proper motion
selection are a number which have parallaxes larger than 7\,mas,
corresponding to distances smaller than 140\,pc. These are clearly not
members of the association. The rest of the discussion will be
focused on the stars located near the Orion molecular clouds, in the
region of the association subgroups.

\begin{figure}[t]
\centerline{\psfig{figure=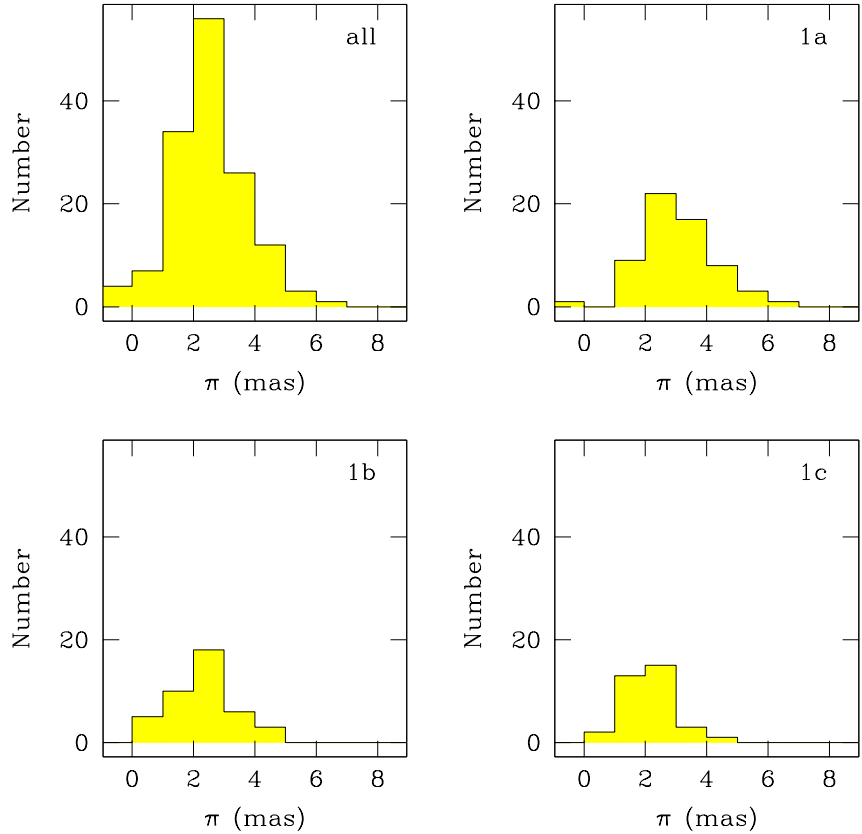,width=12cm}}
\caption{The distribution of parallaxes in the subgroups of Orion
  OB1. The parallaxes are shown for the stars selected according to
  the proper motion criterion given above, and located near the
  positions of the subgroups. Note that the parallaxes in 1a are
  generally larger than those in 1b and 1c, indicating that this
  subgroup is significantly closer to the Sun.}\label{fig-parallaxes}
\end{figure}

Figure~\ref{fig-parallaxes} shows the distribution of Hipparcos
parallaxes for the stars located in the Orion OB1 subgroups and
selected by proper motion according to the criteria above. The stars
were divided among the subgroups following the division given by
Warren \& Hesser (1977a). The parallaxes in subgroup 1a are larger on
average than those in 1b and 1c. This difference is statistically
significant and implies that 1a lies closer to the Sun than 1b and 1c.

From the parallaxes one can infer the distances to the subgroups of
the association. However, the distance depends non-linearly on the
parallax, namely $d=1/\pi_{\rm H}$, where $\pi_{\rm H}$ is the
measured parallax, which is distributed as a Gaussian around the true
parallax $\pi$. In this case the expectation value for $d$,
$E[1/\pi_{\rm H}]$ is in general different from $1/\pi$, even if the
measured parallax is unbiased ($E[\pi_{\rm H}]\approx \pi$). In other
words, $1/\pi_{\rm H}$ is a biased estimate of the star's true
distance. For individual stars the magnitude of the bias depends on
the ratio of the measurement error to the true parallax. The bias can
be calculated analytically as shown by Smith \& Eichhorn (1996).
Another bias in the distance estimate for a sample of stars enters
because of selection effects and completeness limits. An example of
this is the Lutz-Kelker bias (Lutz \& Kelker 1973). The bias due to
selection effects and the completeness of the sample depends very much
on the details of the sample selection. Blindly applying a Lutz-Kelker
correction will lead to erroneous estimates of the bias involved. For
more details and references on how to treat parallaxes correctly see
Brown \etal{} (1997).

In the case of the subgroups of Orion OB1, we proceed as follows. If
one assumes the stars in the subgroup to be distributed in a
spherically symmetric way, and if one assumes that one has sampled the
stellar distribution representatively, then one can show that the
average parallax of that group of stars is an unbiased estimator of
the distance to the centre of that group. Both assumptions have to be
substantiated in the case of Orion OB1. For the more dispersed
subgroups of nearby associations there is no evidence that the
distribution of stars is not spherically symmetric. However, recent
work by Hillenbrand \& Hartmann (1998) on the ONC has shown that
cluster to be elongated. The selection effects that enter into the
Orion OB1 sample are rather complicated. They include both the
selection effects of the 1982 SPECTER proposal as well as the overall
Hipparcos Catalogue selection effects. A specific effort was made,
however, to include as many O and B stars in the Orion region as
possible. Nevertheless, the sample is at least magnitude-limited and
the bias in the calculated distance due to selection effects should be
studied in detail.

Hence, proceeding from the assumptions above, we can calculate the
distances to the subgroups from the average parallax of the members.
For subgroup 1a the average parallax is $3.07\pm 0.15$\,mas (61 stars),
for 1b it is $2.28\pm 0.17$\,mas (42 stars), and for 1c it is
$2.16\pm 0.17$\,mas (34 stars). The quoted errors are the errors in the
mean. The median error on the individual stellar parallaxes is $\sim 1$\,mas 
for all subgroups and the spread in the parallaxes (as measured
by the standard deviation) is of the same order. Hence the depth of
the association is not resolved by the Hipparcos parallaxes. The mean
parallaxes above correspond to distances of 330\,pc for 1a, 440\,pc for
1b and 460\,pc for 1c.

Thus, subgroup 1a is roughly 100\,pc closer to the Sun than subgroups
1b and 1c. The distance to 1a is much smaller than the photometrically
determined distance. This result is also found for many other nearby
OB associations (de\,Zeeuw \etal{} 1997). The smaller distance has
several implications. Subgroup 1a is now located far enough away from
the Orion molecular clouds that it may have triggered star formation
simultaneously throughout the cloud complex, as opposed to triggering
only one part of the cloud. The latter scenario would more likely lead
to a sequence of ages among the subgroups. However, as discussed in the
previous section, such a sequence does not appear to exist, and as
discussed by Hillenbrand (1997), there is a wide distribution of ages
of young stars throughout the molecular cloud complex (outside the
Orion OB1 subgroups). A second consequence of the smaller distance to
1a is that the size, mass, and energy of the Orion/Eridanus bubble
should be lowered by 13\%, 25\% , and 25\% respectively, if one assumes 
that 1a is at the centre of the bubble.
As a consequence the energy requirements for the creation of the
bubble are less stringent. Thirdly, Cunha \& Lambert (1992) noted that
no self enrichment has occurred in subgroup 1b of the association,
whereas it was expected because of supernovae from subgroup 1a.
However, with subgroup 1a located at a large distance from the gas out
of which 1b formed it is very well possible that most of the
supernovae ejecta were lost in the low-density ISM surrounding the
molecular clouds. Hence no self-enrichment of the association occurred
at the location of 1b.

Finally, we would like to stress again that the distance estimates
above are probably biased. A preliminary investigation into this bias
(de\,Zeeuw \etal, in preparation) shows that it is probably small
(around 5\%) and that the distances are underestimated
as derived from the parallaxes. However, examining the parallax
distribution directly (avoiding the bias in the distances) clearly
shows subgroup 1a to be much closer to the Sun than 1b and 1c.

\section{Evidence for Low Mass Stars in OB Associations}
Up to now we have focused on the massive (O and B) stars in Orion OB1.
However, if one extrapolates the IMF for Orion OB1 the bulk of the
stars should be low-mass stars (of mass $\simless$\,2\Msolar). For many
years, the conventional wisdom had been that low-mass stars simply did
not form in regions of high-mass star formation. Regions forming low
mass stars, the T~associations, were dominated by the classical
T\,Tauri stars (cTTS), noteworthy for their strong H$\alpha$~emission,
their UV and IR continuum excesses, and their erratic variability. It
was obvious that there were no high-mass stars in T~associations. For
a field star mass function, the highest mass m$_0$ occurs where the
expectation of N(m$>$m$_0$) is less than unity.  Using the
Miller-Scalo (1979) mass function, there will be about one star of
$>$\,5\Msolar{} for every 100 lower-mass stars. The Taurus-Auriga
T~association, with about 100 cTTS, was not expected to have a
significant population of high-mass stars, and it didn't.

However, this tidy picture began to unravel in the early 1980s.  X-ray
observations with EINSTEIN (and later with ROSAT) revealed a large
number of X-ray sources in these T~associations. Optical follow-ups
(\eg{} Walter \etal{} 1988) showed that these were indeed low-mass PMS
stars, with, in many cases, ages comparable to the cTTS. Walter \etal{}
(1988) concluded that the population of the Tau-Aur association
had been underestimated by nearly a factor of ten. If this were indeed
correct, and the mass function of the association mirrors the field,
then there should be a few tens of B~stars associated with Tau-Aur.

Blaauw (1956, 1984) had identified a loose group of OB stars, the
Cas-Tau group, in this general region of the sky. In fact, subgroups~6
and~7 of this OB group lie at about the distance of the Tau-Aur clouds
and are kinematically indistinguishable from the Tau-Aur T~association. 
These stars are all B5 or earlier. Walter \& Boyd (1991) searched the 
Bright Star Catalog (Hoffleit \& Jaschek 1982) for evidence
of late~B stars which might be associated with this association, and
identified 29 B~stars ($V < \magnit{6}{5}$) which are cospatial with the
Tau-Aur T~association, and which share the space velocity of the T\,Tauri
stars. Walter \& Boyd concluded that the population of the Tau-Aur
association was indeed larger than had been thought, that the mass
function extended up to the early B stars, and was consistent with the
field star mass function\footnote{The B~stars, as well as many of the
  low-mass stars, have ages of up to a few tens of Myr.
  They may be ZAMS, rather than PMS stars. Nonetheless, they are
  kinematic members of the association. As the mass function is
  integrated over time, it is not relevant to argue that the B~stars
  and some of the low-mass stars may represent an earlier episode of
  star formation.}.

If T~associations can contain high-mass stars, can OB associations
contain low-mass stars in the numbers predicted by the field star mass
function? Theory suggested that star formation was bimodal, and that
high- and low-mass stars formed by different processes (\eg{} Larson
1986; Shu \& Lizano 1988). There was no reason to expect 
a universal mass function. Indeed, while there was a decided lack of
evidence for low-mass stars in OB~associations, much of this was due
to observational selection. Low-mass stars are many magnitudes fainter
than high-mass stars, and there were few searches for low-mass stars
in OB~associations\footnote{Historically this was due to the large
  extent of associations on the sky (tens of degrees), which meant
  that they were not amenable to proper motion studies with
  photographic plates. One thus had to rely on large scale meridian
  circle surveys, which generally only included the brighter stars.
  Consequently, membership in OB associations was rather 
  ill-determined for spectral types later than B5.}. However, objective 
prism H$\alpha$ surveys of the Orion association had revealed what
presumably were T\,Tauri stars in the vicinity of $\lambda$\,Ori (Duerr,
Imhoff, \& Lada 1982). The Kiso survey has also found a number of
H$\alpha$-emitting stars throughout the Orion~OB1 association, with
many concentrated towards the belt of Orion (Kogure \etal{} 1989;
Nakano \etal{} 1995).  Furthermore, the concentration of low-mass
stars near the Orion Trapezium has long been known (\eg{} Walker 1969).

Walter \etal{} (1994) used an X-ray-based search to look for the low-mass
population of the Upper~Sco OB~association (de\,Geus \etal{} 1989), and 
found 28 low-mass PMS stars in 7~deg$^2$ of the
association.  After correcting for the incompleteness of the X-ray
sampling, they concluded that the association has a field star mass
function between about 0.2--10\Msolar.  The total number of low-mass 
stars ($<$\,2\Msolar) is about 2000. Part of the reason that
this enormous population of low-mass stars lay undetected is that few
of the stars are cTTS: most have lost their circumstellar disks,
perhaps because of the influence of the winds and ionizing flux of the
B~stars, or because of supernovae in the association.

\section{The Low-Mass Population of Orion OB1}
From the observation that the Upper~Sco OB association has a field
star mass function, it is possible to conclude that there might indeed be a
universal mass function, and if the Orion OB1~association (Blaauw 1991; 
Brown \etal{} 1994) has a similar mass function, there should be 
literally tens of thousands of low-mass stars in
this association. These low-mass stars are important to find and
study, for a number of reasons:
\begin{itemize}
\item The low-mass stars may give a better picture of the history of
  star formation. The locations of low-mass stars in the H-R diagram are
  far more sensitive to age than are the positions of the ZAMS B~stars.
\item The low-mass stars may give a better picture of the kinematics
  of the association. If there is mass-segregation, it will be
  revealed by the low-mass stars. Radial velocities are more easily
  measured for the (narrow-lined) low-mass stars, which have numerous
  sharp lines, than they can be for the B~stars.
\item Most low-mass stars form in OB~associations, not T~associations.
  Studies of stars in T~associations may be misleading, because of the
  external influences of the high-mass star's winds and ionizing
  radiation (not to mention supernovae). These influences may destroy
  accretion disks, altering the mass function, and affecting the
  binary fraction and the formation of planets.
\end{itemize}

\subsection{The Orion Nebula region}
Haro (1953) noted the existence of many emission-line objects in the
vicinity of the Orion Nebula. Walker (1969) obtained spectra of a
number of these stars, and showed that they were members of the ONC
and hence were pre-main sequence stars. Andrews (1981) published a
photometric atlas of Orion showing the existence of many dM stars what
were likely PMS stars. Hillenbrand (1997) has reported on a detailed
photometric and spectroscopic analysis of the low-mass population in
the inner 2.5\,pc of the ONC\@.  There have been spectroscopic studies of
the G~stars in Orion OB1c, just outside the Orion Nebula (\eg{} Smith,
Beckers, \& Barden 1983). Strom \etal{} (1990) have discussed the low-mass
population of the L\,1641 cloud, to the south of the ONC\@. Further 
details can be found in the chapter by Allen \& Hillenbrand.

\begin{figure}[bt]
\centerline{\psfig{figure=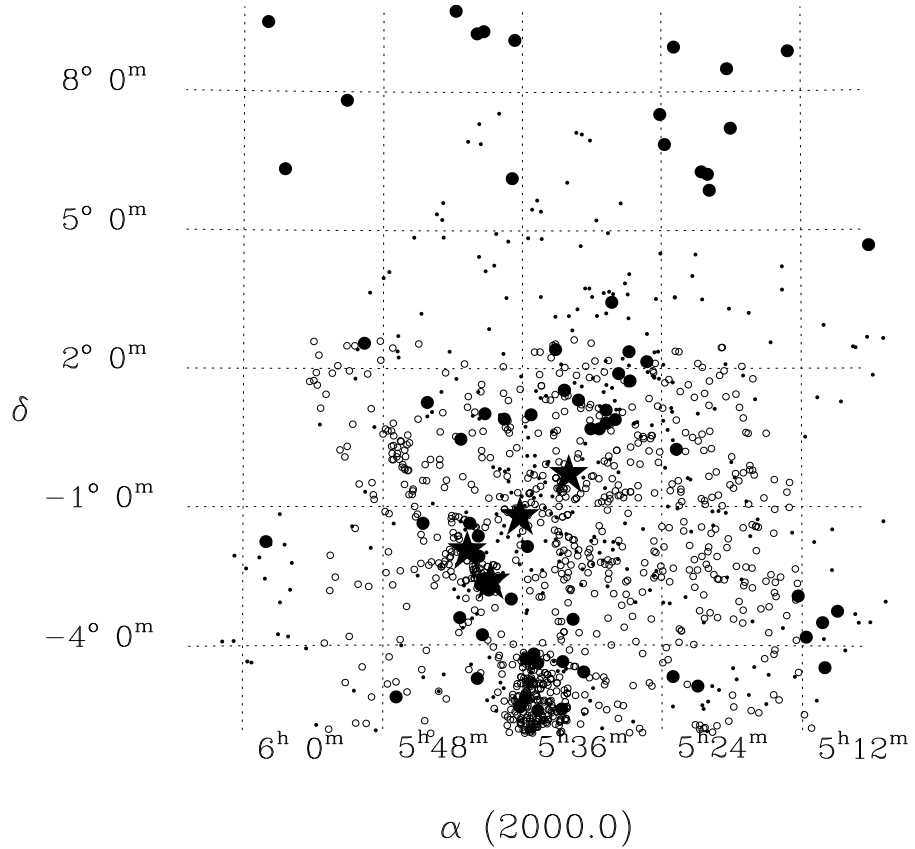,width=12cm,angle=90.}}
\caption{Distribution of X-ray-selected PMS stars and H$\alpha$ sources
  in Orion. The large solid circles are PMS stars identified in the ROSAT
  All Sky survey (Alcal\'a \etal{} 1996).
  H$\alpha$~emitters from the Kiso survey are indicated by
  the open circles. O~and B~star members of the association (Brown \etal{}
  1994) are the small dots. The belt stars ($\delta$, $\epsilon$, $\zeta$)
  and $\sigma$\,Ori are indicated by the large stars.}\label{fig-xdist}
\end{figure}

\subsection{H$\alpha$ surveys}
The Kiso survey (Wiramihardja \etal{} 1989, 1991, 1993; Kogure \etal{}
1989; Nakano \etal{} 1995) is an objective prism survey for
H$\alpha$ emission-line objects in a wide region of Orion (extending
roughly from 5 to 6 hours in Right Ascension, $-10$ to $+5$ degrees in
declination. They identified
about 1200 H$\alpha$ emission-line stars in 300\,deg$^2$. These stars
generally fall in the magnitude range \magap{13}$< V <$\magap{17}. Based on
similarities to the H$\alpha$~emitting population near the ONC, the
authors of this survey conclude that many of these stars are likely to
be T\,Tauri stars. While most of the emission-line stars are
concentrated near the ONC and the L\,1641 cloud, there are many stars in
the region of the belt, with what appear to be significant clumpings
near $\sigma$ and $\zeta$\,Ori (Figure~\ref{fig-xdist}). Kogure \etal{}
(1992) followed-up with low-dispersion spectroscopic observations
of 34 emission line stars in Orion~OB1b, and concluded that they were
indeed T\,Tauri stars based on H-Balmer and \CaII{} K~emission lines.
Nakano \& McGregor (1995) obtained near-IR photometry for a number of
these stars, and concluded that they were indeed mostly T\,Tauri stars.

As the objective prism surveys are only sensitive to stars with strong
H$\alpha$ emission, they miss the naked or weak T\,Tauri stars (nTTS;
wTTS), which greatly outnumber the cTTS (Walter \etal{} 1988;
Neuh\"auser \etal{} 1995). It is not possible to extrapolate from the
emission-line stars to the full number of low-mass PMS stars, but if
the ratio of nTTS to cTTS is 5--10:1 as it is in Tau-Aur, or 30:1 as it
is in Upper~Sco, there may be tens of thousands of low-mass stars
associated with Orion~OB1.

\begin{figure}[t]
\centerline{\psfig{figure=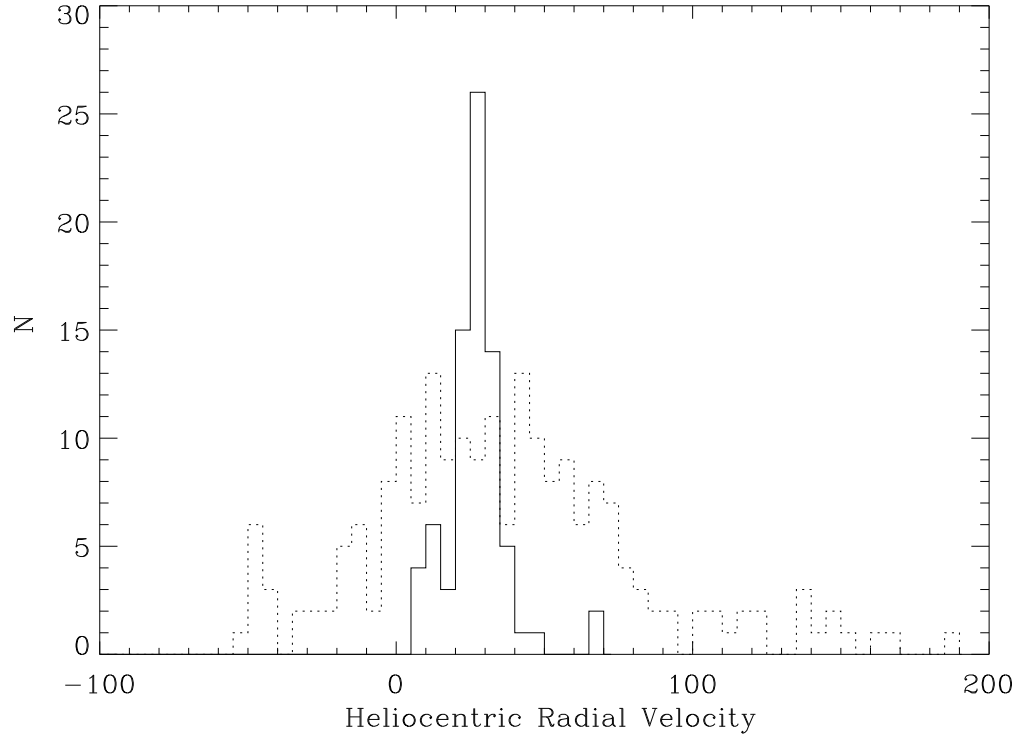,width=\textwidth,angle=90.}}
\caption{The distribution of radial velocities of the stars near
  $\sigma$\,Ori.
  The spectroscopically-identified low-mass PMS stars (solid histogram) are
  well-fit as a Gaussian distribution of mean 25\kmpers{} with
  $\sigma$=5\kmpers. The secondary peak at 12\kmpers{} is due
  to a systematic shift of M~star velocities, and may be an artifact
  of using a sky spectrum as a velocity template. The 
  stars in the sample (dotted histogram) have
  a mean velocity of 31\kmpers{} with
  $\sigma$=37\kmpers.}\label{fig-rvs}
\end{figure}

\subsection{X-ray surveys}

The EINSTEIN observatory was used to observe many regions within the
constellation of Orion. Walter \etal{} (in preparation) have
catalogued over 600 X-ray sources, of which about 200 are low-mass PMS
stars. The space distribution of these PMS stars is shown in
Figure~\ref{fig-xdist}. There are concentrations near $\lambda$\,Ori,
near the ONC and L\,1641, and along the belt, but there are low-mass PMS
stars scattered everywhere within the constellation.

Alcal\'a \etal{} (1996) found a
similar spatial distribution using the ROSAT all-sky survey (RASS).
While the RASS does not go very deep in general, it has the advantage of
offering complete and nearly uniform spatial coverage. Alcal\'a \etal{}
identified 112 new PMS stars in 450\,deg$^2$
(Figure~\ref{fig-xdist})\footnote{The Walter \etal{} and Alcal\'a \etal{}
  surveys are based on fairly low resolution spectra. They did not
  determine radial velocities for the stars. Brice\~{n}o \etal{}
  (1997) argued that many of these stars could be older (100\,Myr),
  foreground ZAMS G~stars, which have not depleted their lithium.
  While this cannot be excluded for the G stars, it cannot be the case
  for the K~and M~stars. The significant enhancement in the density of
  stellar X-ray sources in the direction of Orion (Sterzik \etal{}
  1995) indicates that most of these stars must be associated with
  Orion.}.

Sterzik \etal{} (1995) analyzed the spatial distribution of the RASS
X-ray sources. By selecting sources based on spectral hardness ratios
(X-ray colors), they could select for PMS stars. They showed that the
density of PMS candidate X-ray sources showed significant enhancements
at the locations of Orion~OB1a, OB1b, OB1c, $\lambda$\,Ori, and
NGC\,1788. Further details of these X-ray surveys can be found in
the chapter by Sterzik \etal{} in this volume.

\begin{figure}[t]
\centerline{\psfig{figure=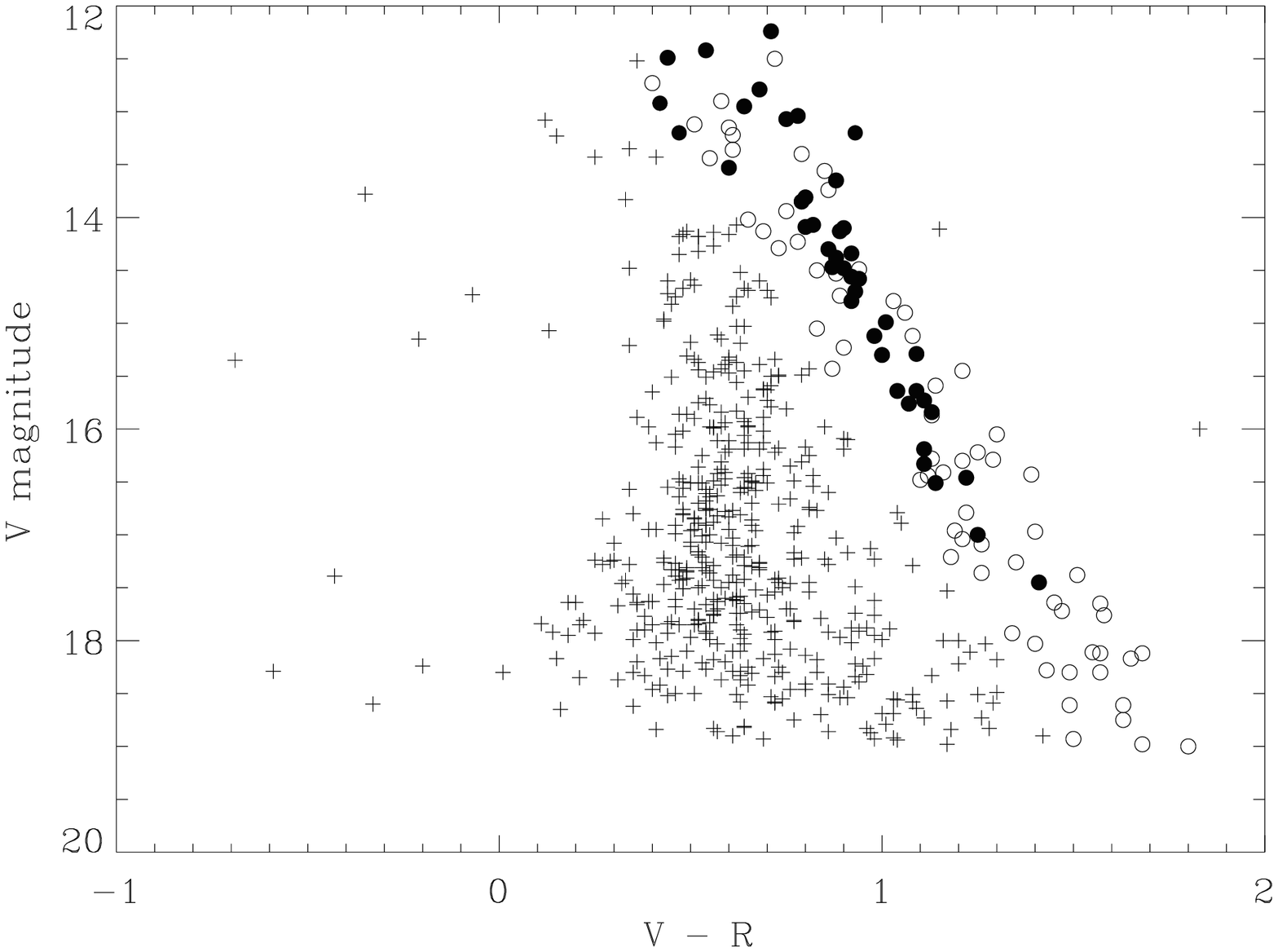,width=\textwidth,angle=90.}}
\caption{The observed color-magnitude diagram for stars with 
  \magap{12}\,$< V <$\,\magap{19} in
  0.15\,deg$^2$ within 15 arcmin of $\sigma$\,Orionis.
  Spectroscopically-identified low-mass pre-main sequence stars,
  marked with solid circles, define the PMS locus. Photometric PMS
  candidates are indicated by the open circles. These data are not
  corrected for extinction; the reddening vector parallels the PMS
  locus.}\label{fig-cmd}
\end{figure}

\subsection{The $\sigma$ Orionis Cluster}

The ROSAT PSPC and HRI observations reveal over 100 X-ray point
sources within 1\degree{} of $\sigma$\,Ori. As part of an investigation of
the low-mass population of Orion, Walter, Wolk, \& Sherry (1998)
investigated the region around $\sigma$\,Ori, a member of Orion~OB1b
and a Trapezium-like system, using multi-object spectroscopy and wide-field
photometry.

Most of the X-ray sources have optical counterparts. Walter \etal{}
observed the optical counterparts of the X-ray sources as well as a
randomly-selected sample of stars in the HST Guide Star Catalog (GSC),
obtaining useful spectra for about 300 stars\footnote{Most of the
  spectra were obtained using the WIYN telescope with the HYDRA multi-object
  spectrometer, as part of the KPNO queue scheduling program.}.  
Among these, they identified 104 likely PMS
stars spectroscopically within 30 arcmin of $\sigma$\,Ori. Primary
identification was made on the basis of a strong Li\,{\footnotesize I}
$\lambda$6707\AA{} absorption line. The H$\alpha$ strengths ranged from an
emission equivalent width of 77\AA{} in a K1 star to apparently normal
photospheric absorption. Radial velocities were determined by
cross-correlating the spectra with spectra of the dusk or dawn sky. At
this fairly low dispersion (1--2\AA{} resolution), uncertainties are
about $\pm$5\kmpers{} for spectra with high signal-to-noise. The distribution
of radial velocities is strongly peaked at the 25\kmpers{} velocity
of the OB association (Figure~\ref{fig-rvs}).

Of the 104 PMS stars, 28 (27\%) are not X-ray sources. This
gives some indication of the completeness of the X-ray sampling.
Some 258 of the optical stars observed were taken from the HST Guide Star
Catalog. These stars constitute a magnitude-limited sample unbiased
with respect to either activity or color.  Of those 258 stars, 57 (22\%)
are likely PMS\@. The estimated space density of PMS stars
with \magap{10}$< V <$\magap{15} within 30 arcmin of $\sigma$\,Ori is 
about 120 stars per deg$^2$.

\begin{figure}[t]
\centerline{\psfig{figure=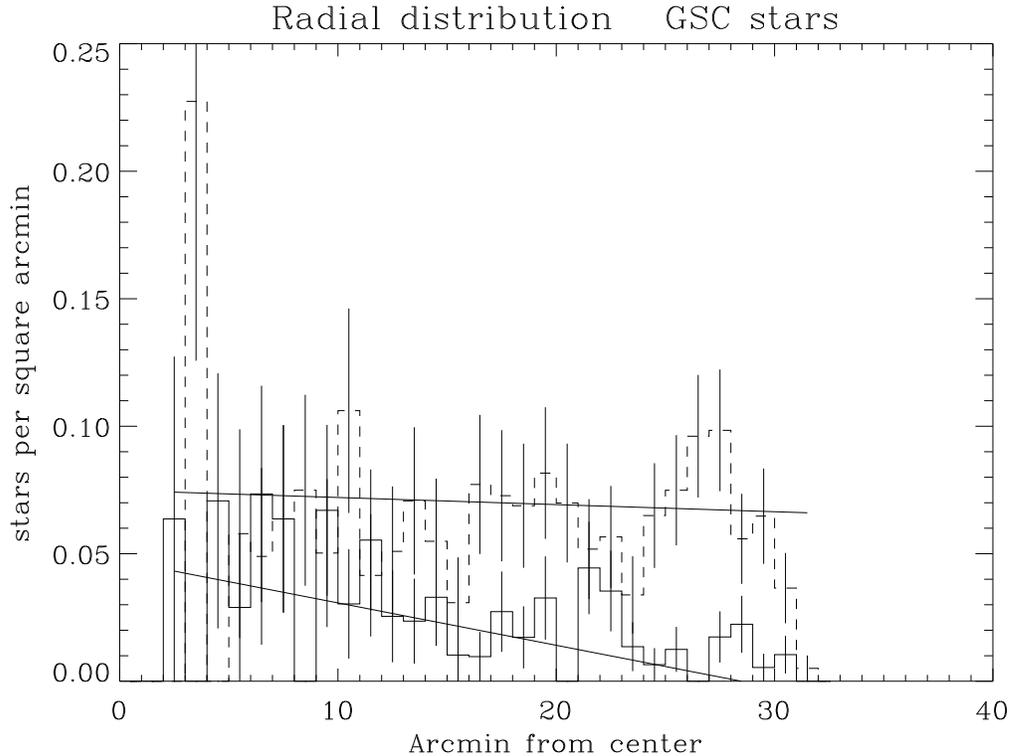,width=\textwidth,angle=90.}}
\caption{The radial distribution of stars in the GSC sample.
  PMS stars (solid histogram) show a clustering towards $\sigma$\,Ori,
  while other stars in the GSC (dashed distribution) do not. The
  probability that the two samples are drawn from the same parent
  distribution is 0.0003. The lines are linear least-squares fits to
  the distribution. The lack of stars within 3 arcmin of $\sigma$\,Ori
  is an artifact of the glare from the 
  \filter{V}=\magnit{3}{8} star.}\label{fig-rad}
\end{figure}

From \filter{UBVRI} images at the CTIO 0.9\,m telescope in January and 
February 1996, Walter \etal{} (1997) sampled stars to \filter{V}=\magap{19} 
in 0.15\,deg$^2$ surrounding $\sigma$\,Ori. The color-magnitude diagram
(Figure~\ref{fig-cmd}) shows a clear PMS locus, well separated from
the background galactic stars. In addition to the 45 PMS stars identified
spectroscopically (which have photometry), there are 65 other stars
along the isochrone to \filter{V}=\magap{19}, most of which are likely 
to be PMS stars.
These 110 photometric PMS stars imply a space density, in the magnitude
range \magap{12}$< V <$\magap{19}, of about 700 per deg$^2$ (or more, if 
many stars are multiple).

The GSC PMS sample shows evidence for clustering. The centroid of the
PMS star distribution is centered on the position of $\sigma$\,Ori.
Summation of the stars into radial bins centered on $\sigma$\,Ori
(Figure~\ref{fig-rad}) shows that the distribution is flat for the
non-PMS stars, but that the radial distribution of the PMS stars is
peaked at $\sigma$\,Ori. The inferred cluster radius (where the density 
of PMS stars reaches zero) is about 0.5\,deg (3.3\,pc).

While the results reported here are still preliminary, with the
analysis continuing, the basic conclusion is secure: there is a
significant population of low-mass PMS stars in this region.  The
stars appear to cluster spatially around $\sigma$\,Ori and the narrowness
of the PMS locus suggests coevality, at the 2\,Myr age of the OB
association. The total inferred mass of this group of stars is
comparable to that of the ONC\@. This $\sigma$\,Ori cluster is 
the {\it second youngest\/} cluster now known after the ONC, and 
may be an evolved analog of the ONC\@.

This results implies that there is substructure in subgroup 1b, and
that the boundaries of the subgroups are not yet well established. As
discussed before, the exact boundaries are important if one wants
interpret the concentration of subgroups in terms of their ages. As
there is no evidence for evolutionary differences between the early-type 
stars in subgroup 1b, it may be that this subgroup formed through
merging of several Trapezium-like clusters that formed at more or less
the same time.

\subsection{Densities of low-mass stars elsewhere in Orion}
Walter \etal{} (unpublished) have also obtained spectroscopic data on five
other regions in Orion~OB1. For each of these fields, they determined
the spectroscopic PMS population (\magap{10}$< V <$\magap{15}), based 
primarily on the Li line strength (photometry exists only for Field~4).  
The space densities of PMS stars are shown in Table~\ref{tbl-sd}.
The space density of PMS stars near $\sigma$\,Ori is not extraordinary.
Field~4, in a nondescript region in Orion~OB1a, has an even higher
space density. The PMS stars in Field~4 have an age of about 10\,Myr,
which is consistent with the age of the OB1a subassociation. This
sample of stars includes no slow rotators (Wolk 1996), and only one
classical T\,Tauri star identified to date.  Note that, in the absence of
photometry and radial velocities, the uncertainties in the densities
in Table~\ref{tbl-sd} are about $\pm$30\%.

\begin{table}[t]
\begin{center}
\begin{tabular}{|lcclc|}
\hline
Field           & RA   & Dec  & Assn.\ & PMS/deg$^2$   \\
\hline
~~4             & 5 24 & $+$1 & OB1a   &   150~~~~     \\
~~6             & 5 31 & $-$1 & OB1a/b &   110~~~~     \\
~~9             & 5 38 & $+$3 & OB1a   &  ~~70~~~~     \\
 15             & 5 31 & $-$3 & OB1c   &  ~~65~~~~     \\
 16             & 5 24 & $-$2 & OB1a   &  ~~40~~~~     \\
$\sigma$\,Ori   & 5 42 & $-$2 & OB1b   &   120~~~~     \\
\hline
\end{tabular}
\end{center}
\caption{Space densities of PMS stars in Orion, 
   \magap{10}$< V <$\magap{15}}\label{tbl-sd}
\end{table}

\section{Conclusions and Future Work}
We have discussed the work done in the past on the early-type stars in
Orion OB1 and described the most recent developments. With the
Hipparcos parallaxes and proper motions now available, our knowledge of
OB associations will be significantly advanced. The most remarkable
result concerning Orion OB1 to come out of the Hipparcos data is the
much smaller distance to subgroup 1a. The derived distances to the
subgroups of Orion OB1 are still preliminary and the biases due to
selection effects should be studied. However, the fact that 1a is much
closer to the Sun than 1b and 1c is a robust result. The implications
are that 1a could have triggered simultaneous star formation
throughout the Orion complex, and the mass and energy of the
Orion/Eridanus Bubble are probably lower than thought previously.

Unfortunately, the Hipparcos proper motions do not allow a clear
kinematic identification of Orion OB1. However, one may attempt to
tackle the problem by using the Hipparcos Intermediate Astrometric
Data. These data essentially allow one to reconstruct the original
observations from which the astrometric parameters of the stars were
derived. One can then try to construct a new solution of the proper
motions and parallaxes of the association stars from these data, by
solving for a common proper motion and parallax assuming all stars
considered are members of the association. For details see Volume~3,
Chapter~17 of the Hipparcos Catalogue (ESA 1997) and van\,Leeuwen \&
Evans (1998). This procedure may provide better insight into the
overall kinematics of the subgroups, as well as into the question of
substructure and subgroup boundaries, and will lead to an improved
estimate of their parallax and the associated errors. Nevertheless, to
really advance our knowledge of the kinematics of the early-type stars
in Orion OB1, accurate ($\sim$\,1--2\kmpers) and homogeneous
radial velocities are needed.

The Hipparcos parallaxes will also be useful in studying the large
scale interstellar medium around Orion OB1. By combining the
parallaxes of stars in the Orion region of the sky with studies of
interstellar lines along the line of sight to these stars, one may be
able to constrain the distances to features in the ISM\@. This will lead
to better insight into the three-dimensional structure of, \eg{} the
Orion/Eridanus Bubble, which will in turn improve the interpretation
of the observations of this interstellar bubble.

We have not addressed the question of the binary population in Orion
OB1. A thorough study thereof would require more accurate knowledge of
the membership of the association than we have at present. For the
early-type stars, a radial velocity survey to search for binaries has
been carried out (Morrell \& Levato 1991), but was limited to the 100
or so brightest stars in Orion OB1. The Hipparcos observations will
provide further information on the binarity of the stars. For the low-mass
stars, radial velocity surveys are also needed to sort out the
frequency of binaries. Accurate knowledge of the binary population is
also needed in order to correctly interpret the kinematics of the
association. The presence of undetected binaries may easily lead to an
inflated velocity dispersion.

The massive stars in OB associations have traditionally received most
of the attention. However, it is now clear that there are significant
numbers of low-mass stars associated with the Orion OB1 association.
Low-mass stars have indeed formed in great numbers in Orion~OB1. The
data are still very incomplete, and the interpretations are still
immature, but we are confident that further study of this low-mass
population will reveal a great deal about the initial mass function in
Orion~OB1, the distribution of stellar ages, and the kinematics of the
association. The best is yet to come.

\end{document}

%% file: macros_ringberg.tex
\newcommand{\degree}{\mbox{$^{\circ}$}}               
\newcommand{\Msolar}{\mbox{\,$M_{\odot}$\/}}          
\newcommand{\HII}{\mbox{H\,{\footnotesize II}}}       
\newcommand{\HI}{\mbox{H\,{\footnotesize I}}}         
\newcommand{\CaII}{\mbox{Ca\,{\footnotesize II}}}     
\newcommand{\magnit}[2]{\mbox{$\mbox{\rm #1}^{\mbox{\rm\footnotesize m}}%
     \!\!\!.\!\,\, \mbox{\rm #2}$}}                   
\newcommand{\magap}[1]{\mbox{$\mbox{\rm #1}^{\mbox{\rm\footnotesize m}}$}} 
\newcommand{\filter}[1]{\mbox{\it #1\/}}              
\newcommand{\oversim}[2]{\protect{\mbox{\lower0.5ex\vbox{%
   \baselineskip=0pt\lineskip=0.2ex
   \ialign{$\mathsurround=0pt #1\hfil##\hfil$\crcr#2\crcr\sim\crcr}}}}} 
\newcommand{\simgreat}{\mbox{$\mathrel{\mathpalette\oversim>}$}} 
\newcommand{\simless} {\mbox{$\mathrel{\mathpalette\oversim<}$}} 
\newcommand{\etal}{\mbox{\hbox{\it et\,al.}}}         
\newcommand{\eg}{\mbox{\hbox{\it e.g.,}}}             
\newcommand{\kmpers}{\mbox{\,km\,s$^{-1}$}}           